\input{aipcheck}

\documentclass[
    ,final            
  ]
  {aipproc}

\layoutstyle{6x9}

\begin{document}

\title{Extracting a free neutron structure function from proton and deuteron deep inelastic scattering data}


\author{O. Hen}{
  address={Tel Aviv University, Tel Aviv 69978, Israel}
}

\author{E. Piasetzky}{
  address={Tel Aviv University, Tel Aviv 69978, Israel}
}

\author{R. Shneor}{
  address={Tel Aviv University, Tel Aviv 69978, Israel}
}

\author{L. B. Weinstein}{
  address={Old Dominion University, Norfolk, Virginia 23529, USA}
}

\author{D. W. Higinbotham}{
  address={Thomas Jefferson National Accelerator Facility, Newport News, Virginia 23606, USA}
}

\begin{abstract}
  Due to the lack of a free neutron target the structure function of
  the neutron cannot be measured directly and is therefore extracted
  from deuteron and proton DIS data. Because the deuteron is a bound
  nuclear system, in order to extract the neutron structure function,
  one needs to apply model dependent theoretical corrections which
  dominate the uncertainty at the large $x_{B}$ region. We present
  here a correlation between the magnitude of the EMC effect and the
  amount of two nucleon Short Range Correlation (2N-SRC) pairs in
  nuclei. Using this correlation we propose a phenomenological
  procedure to extract the free neutron structure function in the
  $x_{B}$ range of 0.3 to 0.7.

\end{abstract}

\maketitle


\section{}

Knowledge of the parton ({\it i.e.,} quark and gluon) distribution
functions of nucleons is of fundamental importance to study the
structure of nucleons, in the search for new physics in collider
experiments, and in analysis of neutrino oscillation experiments
[1]. The $F_{2}$ structure function of the proton and the $u$ Parton
Distribution Function (uPDF) are relatively well constrained from
proton Deep Inelastic Scattering (DIS) over the full Bjorken $x_{B}$
range of 0 to 1. As was discussed by several speakers in this
conference [2,3] the lack of a free neutron target makes the knowledge
of the neutron structure function and the $d$ distribution (dPDF) much
less certain. This is especially clear in the valence quark dominant
(large $x_{B}$) range.

The classic way to extract the neutron structure function and the $d$
quark distribution is by performing DIS measurements on
deuterium. The free neutron structure function is then obtained by
subtracting the known proton structure function from the deuteron
structure function. The dPDF can be extracted from the free proton and
neutron structure functions, assuming  isospin symmetry of the
structure functions. These extractions are done under the assumption
that the deuteron is loosely bound and can therefore be approximated
to a free $np$ (neutron proton) pair. The problem in this procedure is
that the deuteron, even if only loosely bound, is not a free $np$
system. To extract information on the free neutron from the deuteron
one needs to make nuclear corrections that include both ``standard''
effects such as Fermi motion and binding as well as relativistic and
off shell corrections, which depend on the virtuality of the
nucleons. These nuclear corrections and their uncertainty propagate
through the analysis procedure described above and yield a large
uncertainty on the extracted neutron structure function and dPDF,
especially at large $x_{B}$ [4].

We propose here an alternative phenomenological procedure that allow
to extract the free neutron structure function and the dPDF from DIS
off the proton and deuteron without any theoretical calculations. The
procedure is based on a recent report [5] that showed that the
magnitude of the EMC effect, measured in electron DIS, is linearly
related to the Short Range Correlation (SRC) scaling factor, obtained
from electron inclusive scattering at $x_{B}>1$.

\begin{figure}
  \includegraphics[height=.4\textheight]{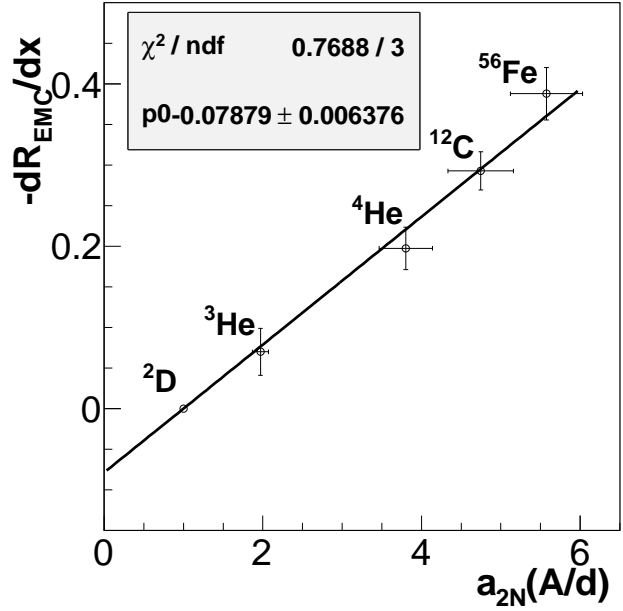}
  \caption{The EMC slopes verses the SRC scale factors. The fit parameter is the intersept of the line. For more details see [5].}
\end{figure}

Figure 1, taken from [5], shows the strength of the EMC effect (defined
by its slope, $dR_{EMC}/dx_{B}$ [6]) versus the SRC scale factors
$a_{2N}(A/d)$ measured for different nuclei [7,8]. As can be seen
clearly in the figure, the EMC strength and the SRC scaling factors
are linearly correlated. Under the assumption that the value
$a_{2N}(A/d)=0$ is the limit of a free $np$ pair with no SRC, one can use this
striking correlation to extrapolate to $a_{2N}(A/d)=0$ and obtain a
phenomenological value to the deuteron to a free np pair DIS cross
section ratio as a function of $x_{B}$ in the range of $0.3\le
x_{B}\le0.7$. Extrapolating the best fit line in Fig. 1 to
$a_{2N}(A/d)$=0 gives an intercept of $\vert dR_{EMC}/dx_{B}\vert=0.079 \pm
0.006$. The difference between this value and the deuteron EMC slope of
0 is a measure of the difference between the deuteron and a free np
pair.  Because the EMC effect is linear for $0.3<x_{B}<0.7$, we can
parametrize the cross section ratio for the deuteron relative to a
free $np$ pair by:
\begin{equation} 
\frac{ \sigma_{d}}{ \sigma_{n} + \sigma_{p}} = 1 - a(x_{B}-b)
\end{equation} 
where $\sigma_{d}$, $\sigma_{p}$, and $\sigma_{n}$ are the DIS
scattering cross sections for scattering off deuterium, the free proton
and the free neutron, respectively. $a=0.079 \pm 0.006$ as extracted from Fig.~1 and
$b=0.31 \pm 0.04$ is the average value of $x_{B}$ where the EMC ratio
is unity as determined from Refs. [6,9], taking into account the
quoted normalization uncertainties.

\begin{figure}
  \includegraphics[height=.4\textheight]{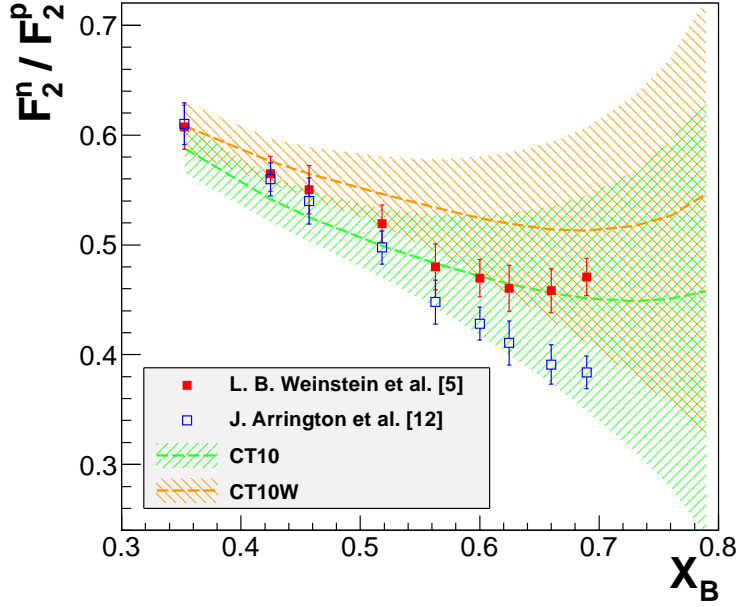}
  \caption{The structure function ratio of the proton and neutron as
    extracted from DIS off the deuteron and proton. The open squares
    (blue online) are from [12] corrected for the nucleon motion
    only. The full squares (red online) include the phenomenological
    correction for nuclear effects as discussed here and in [5]. The
    shaded areas show the structure function ratios calculated using
    CT10 (green online) and CT10W (orange online) PDFs [11]. The width
    of these areas correspond to one standard deviation.}
\end{figure}

If the structure function $F_{2}$ is proportional to the DIS cross
section ({\it i.e.,} the ratio of the longitudinal to transverse cross sections is
the same for the proton, neutron and deuterium, see discussion in [10]),
then the free neutron structure function can be extracted from the
measured deuteron and proton structure functions:
\begin{equation} 
 F^{n}_{2}(x_{B},Q^{2})  =  \frac{2F^{d}_{2}(x_{B},Q^{2})}{1-a(x_{B}-b)} - F^{p}_{2}(x_{B},Q^{2})
\end{equation} 
which leads to:
\begin{equation} 
\frac{F^{n}_{2}(x_{B},Q^{2}) }{F^{p}_{2}(x_{B},Q^{2}) } = \frac{2 \frac{F^{d}_{2}(x_{B},Q^{2}) }{F^{p}_{2}(x_{B},Q^{2}) }}{1-a(x_{B}-b)} - 1
\end{equation}

Figure 2 shows different extractions of ${F^{n}_{2}(x_{B},Q^{2})
}/{F^{p}_{2}(x_{B},Q^{2}) }$ including our extraction using Eq.{} 2
and the world data for the ratio of the deuteron and proton structure
functions, an extraction using the same world data, corrected for the
Fermi motion of nucleons in deuterium alone [12], and two different
PDF analyses of the CTEQ group [11].

The data with the phenomenological correction agrees better with the
CT10W fit which includes $W$ asymmetry and $W$-lepton asymmetry data
[13]. The phenomenological correction which also takes into account
possible modification of the bound nucleon structure functions reduces
the $x_{B}$ dependence of the neutron to proton structure function
ratio at large $x_{B}$.  In collaboration with the CTEQ
group, we plan to continue and examine the effect of the extracted
neutron structure function on the uncertainty in the $d/u$ ratio at
large $x_{B}$.

\begin{theacknowledgments}
  We are grateful for many fruitful discussions with the CTEQ/TEA
  collaborators: J. Huston, H.L. Lai, P. Nadolsky, J. Pumplin,
  and C.P. Yuan and with W. Melnitchouk and A. Accardi at JLab. This work
  was supported by the U.S. Department of Energy, the U.S. National
  Science Foundation, the Israel Science Foundation, and the
  US-Israeli Bi-National Science Foundation. Jefferson Science
  Associates operates the Thomas Jefferson National Accelerator
  Facility under DOE contract DE-AC05-06OR23177.
 \end{theacknowledgments}




\bibliographystyle{aipproc}   

\bibliography{sample}

\IfFileExists{\jobname.bbl}{}
 {\typeout{}
  \typeout{******************************************}
  \typeout{** Please run "bibtex \jobname" to optain}
  \typeout{** the bibliography and then re-run LaTeX}
  \typeout{** twice to fix the references!}
  \typeout{******************************************}
  \typeout{}
 }


\end{document}